# Association of indigo with zeolites for improved colour stabilization

Dejoie C.[a], Martinetto P.[a,*], Dooryhée E.[a,b], Van Elslande E.[c], Blanc S.[d], Bordat P.[d], Brown R.[d], Porcher F.[e,f], Anne M.[a]

[a] Institut Néel, UPR 2940, CNRS, 25 avenue des Martyrs, BP 166, F-38042 Grenoble Cedex 9, France

[b] National Synchrotron Light Source-II, Brookhaven National Laboratory, Upton, NY 11973, USA

[c] Centre de Recherche et de Restauration des Musées de France, CNRS, Palais du Louvre, Porte des Lions, 14 Quai François Mitterrand F-75001 Paris, France

[d] Institut Pluridisciplnaire de Recherche sur l'Environnement et les Matériaux, CNRS, Hélioparc, 2 avenue Pierre Angot, F-64053 Pau Cedex 9, France

[f] Laboratoire de Cristallographie, Résonnance Magnétique et Modélisation, UHP-CNRS, Faculté des Sciences BP 70239 , F- 54506 Vandoeuvre-les-Nancy, France

[g] Laboratoire Léon Brillouin, CEA-CNRS, F-91191 Gif-sur-Yvette cedex, France

* Corresponding author: pauline.martinetto@grenoble.cnrs.fr

Tel : +33(4)76887414

Fax : +33(4)76881038

**Abstract**

The durability of an organic colour and its resistance against external chemical agents and exposure to light can be significantly enhanced by hybridizing the natural dye with a mineral. In search for stable natural pigments, the present work focuses on the association of indigo blue with several zeolitic matrices (LTA zeolite, mordenite, MFI zeolite). The manufacturing of the hybrid pigment is tested under varying oxidising conditions, using Raman and UV-visible spectrometric techniques. Blending indigo with MFI is shown to yield the most stable composite in all of our artificial indigo pigments. In absence of defects and substituted cations such as aluminum in the framework of the MFI zeolite matrix, we show that matching the pore size with the dimensions of the guest indigo molecule is the key factor. The evidence for the high colour stability of indigo@MFI opens a new path for modeling the stability of indigo in various alumino-silicate substrates such as in the historical Maya Blue pigment.

Keywords : **indigo, zeolite, pigment, Raman, reflectance**

## INTRODUCTION

Nanocomposite materials, at the interface of the organic and inorganic realms, offer a wide range of tailor-made functional materials in terms of chemical and physical properties.[1] Hybrid materials, combining the properties of an inorganic host and the tailorable properties of organic dyes, are of wide interest for paint and pigment industry. There is thus current interest in hybrid pigments for cosmetics,[2] phototherapy[3] and paints.[4] Interestingly, the mixing of inorganic and organic components can also be found in older contexts.[5] The quest for durable dyes led several ancient civilizations to the manufacturing of artificial pigments, such as lacquer pigments[6] or Maya Blue (MB).[7,8] MB is formed by heating a mixture of a fibre clay and indigofera leaves, and was extensively used in Mesoamerica (300-1500 AD), on frescoes, potteries, sculptures and ritual objects. Indigo molecules are generally thought to be intercalated in tube-like channels of palygorskite or sepiolite clays.[9,10] MB is a robust pigment, capable of resisting biodegradation,[11] and it has survived for centuries in the Meso-American tropical forest.

The ageing, non-toxicity and durability of such historical compounds have uniquely been attested over a time scale and in environments which cannot be normally achieved in the laboratory. Our approach is to investigate these ancient, archaeological remains and to apply this knowledge in the design of new, archaeo-inspired hybrid pigments.[12] An important advantage of MB-like composites is the protection of the dye molecules against chemical attack, photo-bleaching or thermal decomposition. Light sensitive 1,6-diphenylhexatriene (DPH) for example is considerably stabilized when inserted into zeolite L.[13] Inclusion of $\alpha,\omega$-diphenylallyl cations (DPP) into ZSM-5 is observed to lead to a major increase in their stability against reaction with $H_2O$.[14]

A number of studies already deals with MB-like dyes. In recent papers, Lima et al.[15,16] attempted to copy the principles of colour fixation inherent to the Maya Blue. These authors have tested the stabilization of some betalain extract from the Mexican bougainvilleas in 3 distinct hosts: layered double hydroxides (Mg,Zn–Al–O LDH's), gamma alumina and zeolites. Marangoni et al.[17] incorporated several blue dye molecules (Evans blue, Chicago sky blue, Niagara blue) within the galleries of negatively charge layered double hydroxide (LDH), before dispersion into polystyrene. Hybrid filler gives rise to blue coloured plastic films, reminiscent of the blue Maya effect. The Maya Blue pigment also inspired Zhang et al.[18] in the elaboration of a photocatalyst for a novel efficient hydrogen evolution system, in which palygorskite clay acts as a matrix and Eosin Y as a photosensitizer. Very recently, we introduced a new Maya Blue-inspired hybrid pigment, with the incorporation of indigo molecules within the channels of the MFI zeolite.[12] Combining optical spectroscopies (UV-Visible reflectance and fluorescence) and X-ray diffraction techniques, we showed that indigo diffuses as monomers through the channel network of the zeolite and sits at particular crystallographic positions while the matrix is maintained in a metastable state. This new hybrid pigment presents a high thermal stability, and resists visible light irradiation.

The objective of this paper is to investigate the colour stability of the hybrid pigments elaborated by insertion of the indigo dye in several zeolite matrices. Indigo dye, used by ancient Mayas to form archaeological Maya Blue, is a pigment well known by ancient civilizations.[19] The deep blue colour of the dye is attributed to the presence of the NH donor groups and of the C=O acceptor groups substituting the central double bond C=C of the molecule.[20] Indigo is particularly unstable in oxidising conditions. In nitric acid media, the central double bond is broken and indigo transforms itself into yellow isatin (Fig. 1). This property is used in this paper as a test to determine the stability of the freshly synthesized indigo@zeolite hybrids.

In this work, we elaborate different hybrid systems by mixing indigo powder with LTA zeolite, Mordenite, and MFI zeolite (Table 1). These particular zeolites have been chosen because of i) their channel/cage dimensions and the possibility to obtain a surface or bulk hybrid; ii) their chemical composition and the possibility of interaction with Al atoms. Colour stability after the oxidising test is investigated by colorimetry and UV-Visible reflectance; chemical evolution of the organic dye is followed by Raman spectroscopy. We demonstrate that colour stability is governed more by the channel/cage dimensions of the zeolites, than by their chemical composition or by the occurrence probability of organic/inorganic interactions.

**EXPERIMENTAL**

**Preparation of the hybrid pigments and oxidising test**

MFI zeolite samples were prepared at the Laboratoire de Cristallographie, Résonnance Magnétique et Modélisation, Nancy, using an adaptation of the fluoride route to hydrothermal synthesis[21] with tetrapropylammonium cation as templating agent. Pure $SiO_2$ (silicalite) material and $Si_{1-x}Al_xO_2$ x~0.025 (ZSM-5) were grown and their chemical composition checked by EPMA-WDS (Electron probe micro-analyser – Wavelength-dispersive spectroscopy). Samples were calcined at 600°C in order to remove the organic template. Mordenite powder (MOR) was crystallized according to the protocol of Sano et al.,[22] with addition of 1-butanol (BuOH) to enhance the crystallinity.[23] LTA zeolite was purchased from E.J. Lima Munoz (Mexico) and chemical composition was checked by SEM-EDX (Scanning electron microscopy – Energy-dispersive spectroscopy).

Dehydroindigo was synthesized using Kalb's synthesis procedure.[24] Purity of the organic compound was checked by infrared spectroscopy with the extinction of the NH vibrational band of indigo.

Synthetic indigo (Sigma-Aldrich) was used in the formation of all the hybrid pigments. A solid/solid phase reaction procedure was employed. Powders of inorganic matrices (grain size ~2μm) and indigo (1%wt.) were finely hand-ground and mixed in a mortar. The resulting powder was pressed into a pellet to enhance contact between the two components. The pellets were placed in an oven and heated for 5 hours in air at 250°C. After the heating phase, pellets were re-ground and washed with acetone to remove unreacted indigo.

Colour stability of the samples was tested in nitric acid conditions at room temperature. Hybrid pigments were stirred in a $HNO_3$ solution for different duration times and different concentrations (Table 1, Table 2).

**Techniques**

Diffuse Reflectance UV-Visible spectra were recorded on a Varian Cary 5 spectrometer (Institut Pluridisciplnaire de Recherche sur l'Environnement et les Matériaux, Pau) equipped with an integrating sphere. Samples were pressed between two quartz windows in a home-made sample holder. The spectra are displayed as $F(R)$ Kubelka–Munk units, with: $F(R)=(1-R)^2/2R=k/S=\varepsilon c/S$ where $R$ is the corrected reflectance, $S$ stands for the scattering coefficient (depending on the size and form of the particles), $\varepsilon$ the molar absorption coefficient of the analyte and $c$ its molar concentration. The Kubelka–Munk transformation thus converts a reflectance spectrum into a spectrum similar to conventional absorbance for solution samples.

Colorimetric measurements were performed on a RUBY spectrocolorimeter (STIL) at the Centre de Recherche et de Restauration des Musées de France, Paris, equipped with a backscattering geometry.[25] Data were collected on the powder samples placed on a slide glass, using a 4mm spot. CIELAB La*b* (L=lightness; a* = from green to red; b* = from blue to yellow) colour space coordinates were calculated using the D65 illuminant.[26]

The Raman spectra were recorded on a LABRAM Jobin-Yvon spectrometer at the Laboratoire des Matériaux et du Génie Physique, Grenoble, using a Ar+ ion laser (488.0 nm) as excitation line and a laser power of ~100µW at the sample. Calibration was achieved using a silicon standard. Some milligrams of hybrid samples were deposed on a glass slide under the microscope. Accumulation of 3 scans, 200 seconds exposure was recorded using a ×50 objective for each spectrum. Possible irradiation damages were investigated tuning the incident laser power and time of exposure. No evidence of sample damage was noticed for the indigo@silicalite hybrids studied by Raman spectroscopy. The spectrometer was controlled with the LabSpec software (Jobin-Yvon Horiba).

The Infra-red spectra were recorded on a FT-IR NEXUS microscope, at the ID21 beamline at the ESRF, Grenoble using the conventional thermal source. Samples were mixed with KBr, and pressed into a pellet. Spectra were recorded between 400 and 4000 $cm^{-1}$.

**Data processing**

Principal Component Analysis (PCA) was applied to experimental Raman spectra in order to visualize the progressive transformation of the indigo dye occurring in different nitric acid conditions. Prior to PCA, Raman spectra were submitted to data pre-treatment. Spectra were smoothed using a Savitsky-Golay 5 points process, followed by first derivative to reduce baseline-offset effects (Origin v5.3 software) and normalisation. Spectral domain was reduced to 1200-1750$cm^{-1}$. PCA is a procedure employed to reduce the dimensionality of a dataset and to reveal the variance among multivariate data. The raw data X-matrix (19 Raman spectra) is decomposed into a sum of two smaller matrices : $X=TP^T+E$, where T is the score matrix containing information about the objects, P the loading matrix containing information about the variables, and E a residual matrix. Uncorrelated loading vectors are a linear combination of the original variables. A new space is thus generated on which the objects and the variables are projected. Calculations were performed using the Statistica v7.1 software (www.statsoft.com).

## RESULTS

### Colour stability of indigo@zeolite complexes

The raw inorganic matrices do not show any reflectance band in the 400-800nm range. Consequently, the resulting reflectance spectra and colorimetric coordinates only depend on the electronic state of the organic guest molecule.

UV-Visible reflectance spectra for the three indigo@zeolite systems before heating, after heating, and after the oxidising test (10 minutes in concentrated nitric acid) are shown in Fig. 2. Before heating, a large band in the 400-800nm range is observed for the three systems (Fig. 2a, 2b and 2c). As observed when diluted with an inert matrix like KBr,[27] this broad band is attributed to powdered indigo. The heating phase does not really affect the indigo@LTA reflectance spectrum (Fig. 2d). Despite supplementary absorption bands in the 400-500nm range, reflectance maximum of the indigo@MOR heated sample is still found around 660nm (Fig. 2e). The indigo@MFI sample presents different features, with the maximum reflectance blue shifting from 660nm to 615nm after the heating process (Fig. 2f). This shift is attributed to the diffusion of indigo monomers inside the zeolite channels.[12] No band is observed anymore for the indigo@LTA sample after the oxidising test (Fig. 2g). A 415nm band perdures for indigo@MOR (Fig. 2h), which corresponds to the value found for diluted yellow isatin in benzene (404nm) and acetonitrile (414nm).[28] The oxidising test performed on the indigo@MFI system does not provoke any radical change as opposed to the two first samples. Only a shift of the maximum reflectance band from 615nm to 590nm is noted (Fig. 2i).

The results of the oxidising test carried out on the three indigo@zeolite samples and their associated colorimetric coordinates in the La*b* system are given in Table 1. Projection in the a*b* space of the colorimetric measurements is presented in Fig. 3. Before oxidation, all the hybrids are situated in the blue zone. After oxidation, the colorimetric coordinates of indigo@MOR and indigo@LTA reveal a displacement towards a more yellow/red zone. The high value of the L coordinate for the indigo@LTA sample (L=90.2) is characteristic of a drastic increase of the white component and the disappearance of the colour. For the indigo@MFI hybrid, a red colour shift is observed with the increase of the a* value from -2.64 to 4.67.

The reflectance spectra after the oxidising test are in agreement with the colour measured: a white sample using a LTA matrix (no reflectance band anymore), a yellow compound using mordenite, and a blue/violet one with the MFI zeolite. Similar results are obtained using different MFI types: silicalite or ZSM-5. Following the colour persistence criteria in oxidising conditions, only the indigo@MFI sample successfully passed the test (Table 1). The two other systems based on LTA zeolite and mordenite fail, and hence will not be considered any longer. The possible chemical evolution of the organic dye will be investigated on samples using only the silicalite matrix.

**Indigo@silicalite hybrid: evolution of the organic molecule**

*Raman spectrum of the indigo@silicalite hybrid*

The Raman spectrum of the indigo@silicalite (MFI) hybrid is shown in Fig. 4. For comparison, the spectrum of indigo powder has been added. Contribution of the inorganic zeolite under similar experimental condition is negligible in the 400-1800cm$^{-1}$ range and the Raman signal only corresponds to that of the organic molecule. The main Raman bands for the indigo@silicalite system are found at 543, 597, 671, 940, 1014, 1097, 1144, 1250, 1311, 1360, 1382, 1462, 1486, 1589, 1631 and 1701 cm$^{-1}$. The absence of defects and of substituted aluminum atoms in the framework of the silicalite prevents the formation of strong organic-inorganic interaction such as hydrogen bonds. Indigo molecules can thus be considered as isolated monomers "diluted" in the channel network of the zeolite. The significant differences noted with the Raman signal of indigo powder[29,30] are imparted to the presence of monomers of indigo inside the zeolite channels, the disappearance of the hydrogen intermolecular bonding responsible for the cohesion of the crystal,[12] and the absence of new bonds between the molecule and the inorganic matrix. To our knowledge, very few studies present the Raman features of indigo in a monomeric state, due to the very low solubility of the dye.[31] On the other hand, the Raman signature of indigo associated with inorganic matrices has been previously reported[32]. Coupry et al.[33] investigated "jean" clothes and attributed the apparition of the $B_u$ symmetry modes to a change in planarity of the molecule. Indigo associated with the palygorskite clay was also extensively studied.[30,34,35,36] Strong interactions via Al-N bonds between the amine groups and the aluminum atoms of the palygorskite,[37] or hydrogen bonds between the amine and carbonyl groups of indigo and structural water of the clay[35] have been found. In the latter case, although indigo is considered as a monomer,[38,39] Raman features are strongly dependant on the interaction involved. In our case, silicalite can be considered as a "solid solvent" for indigo, and this enables to obtain for the first time spectroscopic information on indigo single molecules without any solvent contribution or organic-inorganic interaction dependence.

*Indigo@silicalite hybrid under oxidising conditions*

We extended the oxidising tests on the indigo@silicalite system by varying the nitric acid concentration and the duration time of the test. The experimental conditions and the resulting colorimetric coordinates can be found in Table 2. The corresponding projection on the a*b* space is presented in Fig. 5. For concentrated nitric acid (ox, 14mol/L), the colour shifts from blue (as-prepared, La*b*=43.11,-2.64,-30,22) to a more violet colour (25 hours,

La*b*=33.60,15.10,19,39) with a significant increase of the a* value. For a 3 times diluted concentration (dil3), a similar effect is observed with a lower magnitude. At this concentration, a 25 hours duration time is required to really affect the colour of the complex with the gradual occurrence of a red component. For a more diluted concentration (dil6), the only change occurs in the b* value which contributes to slightly modify the blue hue of the hybrid.

In order to follow the possible evolution of the organic molecule in oxidising conditions, Raman spectra have been recorded on the samples submitted to the various nitric acid tests (Table 2). We observed different kinds of transformation on the organic molecule depending on the concentration and the duration of the oxidising treatment. The two characteristic spectra are presented on Fig. 4. After 21 days in a dil6 nitric acid solution, the Raman bands attributed to indigo inside the zeolite have almost completely disappeared whereas new bands at 1155, 1160, 1382, 1447 and 1541 cm$^{-1}$ increase. This new form is called the A form. A more concentrated treatment (dil3 or ox) has other consequences on the Raman response. The corresponding Raman bands peak at 658, 1305, 1360, 1588, 1639 and 1714 cm$^{-1}$ (B form).

In order to better understand the progressive transformation of the indigo molecule into these two new forms, the experimental Raman spectra were analyzed using a Principal Component Analysis. This data reduction method has been previously used as spectral searching algorithm for pigment identification[40-42]. Calculation of the principal components was performed using 19 spectra corresponding to the 3 different forms obtained (the indigo form prior to any oxidising test (6 measured points), the A form (7 measured points), and the B form (6 measured points). All the other intermediate points are introduced as supplementary cases in order to be able to compare their relative positions in the new space created. More than 90% of the spectral variance of the data set is explained by the first six principal components (not shown). According to the loading plot (Fig. 6), the first principal component (PC1) is positively correlated with the 1250, 1381, 1631, 1701 cm$^{-1}$ bands and negatively correlated with the 1598 and 1711 cm$^{-1}$ bands, which can be related to the progressive transformation into the B form. The second component (PC2) is mainly negatively correlated to the 1379, 1450 and 1533cm$^{-1}$ bands, corresponding to the apparition of the A form Raman vibrational modes. These two first principal components account for the largest spectral variation, with a similar weight. The score plot of the two first principal components is shown on Fig. 7, illustrating the evolution of the spectra as experimental oxidising conditions change. We can clearly observe the presence of two "final" groups depending on the intensity of the oxidising treatment, separated from a third group corresponding to the initial indigo form. The different points for a particular sample give an idea of the heterogeneity of the measurements.

Samples after a soft oxidising attack (dil6) are negatively attracted by PC2, and a progressive shift is observed towards negative PC2 on increasing the duration of the treatment from 0 day to 21 days. Transformation into the A form is thus progressive and requires time to be effective. After a more concentrated nitric acid attack (dil3 and ox), scores are progressively negatively shifted towards the PC1 axis. One can note that a 25 hours treatment in a 3 times diluted nitric acid media (dil3-25h) appears to have a similar effect than 10 minutes in a concentrated medium (ox-10m). Emergence of the B form is thus delayed when using a dil3 nitric acid concentration, as proved by the intermediate position of the dil3-2h group. Comparing the relative position of the ox-2h and the ox-25h groups, complete transformation into the B form is achieved after 2 hours using concentrated nitric acid.

**DISCUSSION**

Among the three indigo@zeolite systems tested in this study, only the indigo@MFI hybrid presents a conclusive colour stability under oxidising condition. The presence of aluminum atoms in the zeolite framework does not constitute a key factor to obtain a stable pigment. LTA zeolite and Mordenite (MOR) are Al-rich and possible bonding between the organic molecule and the inorganic matrix could be expected to form as found in some lacquer pigments.[6] Bonds, if they ever form, are not efficient enough to produce a durable compound and to prevent the destruction of the indigo molecules in oxidising conditions. This result is confirmed by the synthesis of a stable hybrid using the high silica MFI zeolite (silicalite).

Channel cross section or cage dimensions with respect to the size of the indigo molecule are of prime importance. The breaking of the central alkene function in presence of nitric acid requires room for a perpendicular bridging intermediate (Fig. 8).[43] Regarding the dimension of the cage entrance of the LTA zeolite (4.1*4.1 Å²), indigo molecule (4.8*12.3 Å²) cannot enter this zeolite. In oxidising condition, indigo transforms itself into isatin on the surface with no steric constraint, and the resulting compound has the white colour of the raw LTA zeolite. Considering the dimensions of the mordenite (7*6.5 Å²) and that of the MFI zeolite (5.1*5.5 Å²), indigo is able to enter the channels of both these matrices during the synthesis process. The persistence of a reflectance band (415nm with MOR and 600nm with MFI, Fig. 2) after the oxidising test confirms the presence of internal molecules. The yellow compound obtained after the oxidising test with the mordenite indicates that transformation of indigo into isatin occurs inside the channels. The large channel structure of this zeolite does not prevent the perpendicular bridging intermediate to form, and indigo transforms into isatin. On the contrary, MFI zeolite fits better the indigo molecule, and the reaction intermediate cannot be accommodated so easily, preventing total oxidization of the organic dye and colour extinction.

Depending on the oxidising test (acid concentration and duration time), colour of the indigo@MFI system varies from blue to more violet. More precise tests enable us to follow the chemical evolutions of the organic dye associated with this colour change. Formation of the A form occurs in soft oxidising conditions (Fig. 5 and 7). Under specific conditions,[24] indigo is able to form an intermediate oxidising form called dehydroindigo (Fig. 1). In order to check the possibility of the transformation of indigo into dehydroindigo in the MFI zeolite, dehydroindigo was synthesized as described in the experimental section. Raman spectrum of freshly synthesized dehydroindigo is presented in Fig. 9. Raman bands are found at 1158, 1165, 1386, 1452, et 1533 cm$^{-1}$, and fit well with those of the A form. The shift observed for some band positions just as the differences in some of the Raman band intensities can be due to the difference between the powdered form for the dehydroindigo reference spectrum and the single molecule form in the MFI zeolite channels. Formation of this intermediate oxidising form does not really affect the original blue colour of the complexes (Fig. 5).

Concerning the second molecular change occurring under concentrated nitric acid conditions (Fig. 5 and 7), complementary IR experiment was performed to obtain other evidences on the B form. On the major part of the spectrum, vibrational modes of the organic molecule are partially occulted by the IR signal of the inorganic MFI zeolite. Nevertheless, by magnifying the 1250-1800cm$^{-1}$ range, signature of the organic component is obtained with, in particular, the emergence of two bands at 1542 and 1346 cm$^{-1}$ (Fig. 10). Presence of the nitro group (-NO$_2$) in some benzene derivative gives two bands in the 1510-1580cm$^{-1}$ and in the 1325-1365 cm$^{-1}$ ranges.[44] These new bands could correspond to respectively the asymmetric and the symmetric stretching modes of -NO$_2$ group in substitution on the benzene rings of the indigo molecule. Formation of this nitro-compound is associated with a colour change from blue to violet. The presence of an electron attractor group such as NO$_2$ on the indigo molecule could be coherent with the 615nm to 590nm absorption shift and this associated colour change when forming the B form (Fig. 5). The colour modification is progressive when using a diluted 3 times nitric acid medium (Fig. 7), and the violet colour is only effective after a 25 hours treatment. On the contrary, only a 10 minutes treatment is needed to obtain the violet hue when using concentrated nitric acid.

**CONCLUDING REMARKS**

In this study, indigo molecules are compounded with three zeolites, LTA zeolite, mordenite and MFI zeolite. Only the indigo@MFI zeolite sample passes the colour stability test in oxidising conditions. Protection of the indigo molecule by the MFI zeolite is attested by the persistence of the blue/violet colour. The indigo molecule can be slightly modified, depending on the intensity and the duration of the oxidising test. Raman signal of the MFI zeolite being negligible in the considered range, we show that Raman spectroscopy is an efficient probe to follow changes of the organic part. Formation of an intermediate oxidized form (dehydroindigo) and of a nitro-compound was thus revealed. Complementary investigations combined with molecular simulation are required to establish the exact reaction occurring in the channels of the MFI zeolite, which results in the transformation of the indigo molecule into dehydroindigo and nitro-indigo. However, destruction of the molecule and formation of isatin after breaking of the central double bond is never observed with the MFI zeolite. As indigo stability in presence of nitric acid mainly depends on steric considerations, the main criterion to obtain a stable indigoid pigment is found to be the opening dimensions of the inorganic matrix, rather than its chemical composition.

**ACKNOWLEDGEMENTS**

Colorimetric measurements at the C2RMF Paris benefited from the support of J-J. Ezrati. Marine Cotte provided assistance for the infrared studies at the ID21-ESRF beamline

(Grenoble). J. Kreisel and O. Chaix (LMGP Grenoble) helped with the Raman analyses and Ph de Parseval (LMTG Toulouse) provided support for the microprobe analyses. C.D. acknowledges the CIBLE and MACODEV grants from Région Rhône-Alpes.

**TABLE CAPTION**

**Table 1** - Colour stability tests (10 minutes in concentrated nitric acid) for the different hybrid systems (1%wt. indigo) - Colorimetric indices are given in the La*b* space – LTA = indigo@LTA zeolite; MOR = indigo@mordenite; MFI = indigo@silicalite

| Label | Zeolite name and theoretical formula | Channel/cage dimensions[45] | L a* b* Before test | L a* b* After test | Resulting sample colour | Stability |
|---|---|---|---|---|---|---|
| MOR | Mordenite $[Na_8(H_2O)_{24}][Si_{40}Al_8O_{96}]$ | 1-D channel 7.0x6.5 Å² | 27.63 -2.49 -15.89 | 30.43 2.36 2.07 | Yellow | NO |
| MFI | Silicalite $(SiO_2)_{96}$ | 2-D channel Straight: 5.3x5.6 Å² | 43.11 -2.64 -30.22 | 31.79 4.67 -13.25 | Blue/Violet | YES |
| LTA | LTA zeolite $Na_{96}[(AlO_2)_{48}(SiO_2)_{48}].215H_2O$ | Cage 4.1x4.1 Å² | 44.91 -0.11 -14.26 | 90.2 1.01 1.20 | White | NO |

**Table 2** - Oxidising tests for the indigo@silicalite (MFI) sample in presence of $HNO_3$ - Colorimetric indices are given in the La*b* space

| | 0 min | 10 min | 2h | 25h | 21days |
|---|---|---|---|---|---|
| 14mol/l (-ox) | 43.11 -2.64 -30.22 | 31.79 1.67 -13,25 | 28.64 3.91 -12.72 | 33.60 15.10 -19.39 | - |
| 5mol/l (dil3) | | 34.47 -2.06 -19.55 | 29.63 -0.66 -14.81 | 28.78 0.78 -9.14 | 43.72 3.62 -17.82 |
| 2.5mol/l (dil6) | | 42.19 -2.81 -29.12 | 28.38 -2.45 -16.64 | 29.73 -0.70 -13.03 | 38.72 -0.90 -12.15 |

**FIGURE CAPTION**

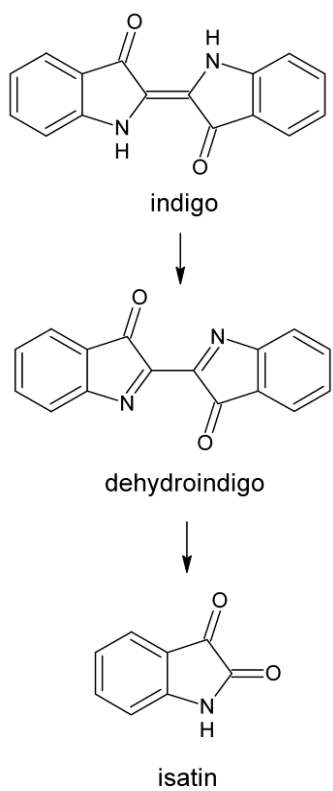

**Fig. 1 -** Oxidisation of the indigo molecule

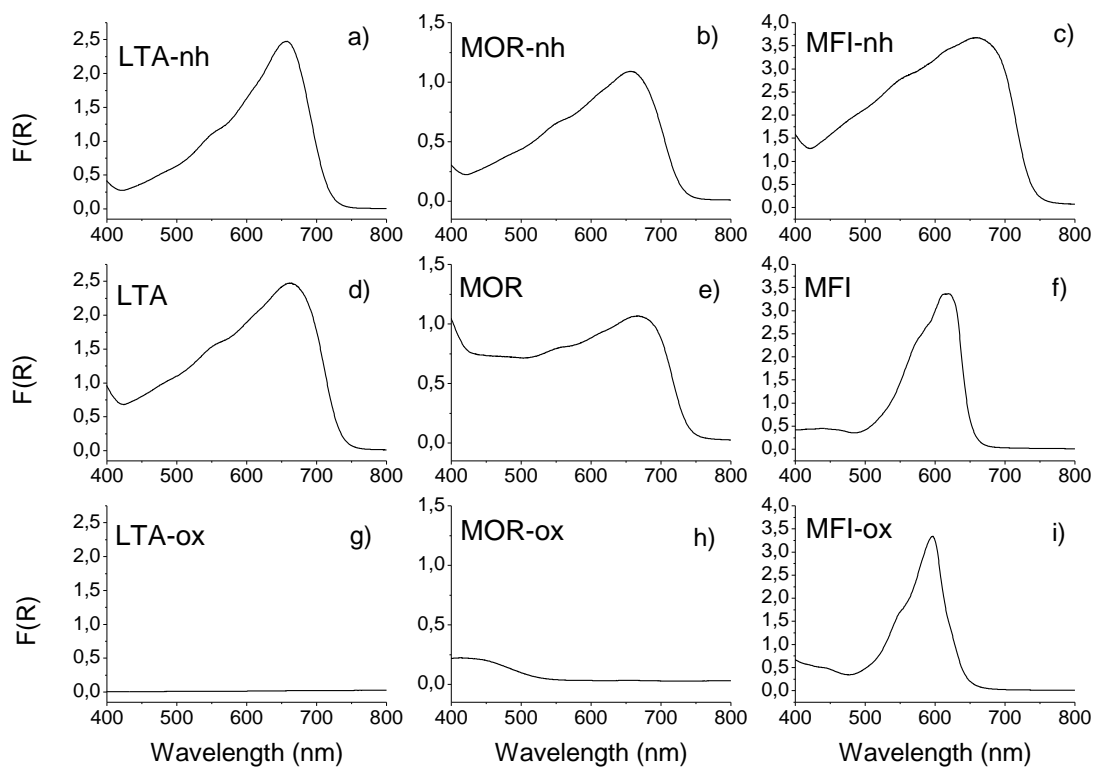

**Fig. 2** – Reflectance spectra of 3 indigo@zeolite samples (LTA = indigo@LTA zeolite; MOR = indigo@mordenite; MFI = indigo@silicalite), before the heating phase (a), b) and c)), after the heating phase (d), e) and f)), and after the oxidising test (10 minutes in concentrated $HNO_3$, 14mol/l, g), h) and i)). Suffixes –nh and –ox respectively refer to "non-heated" samples and samples after the oxidising test.

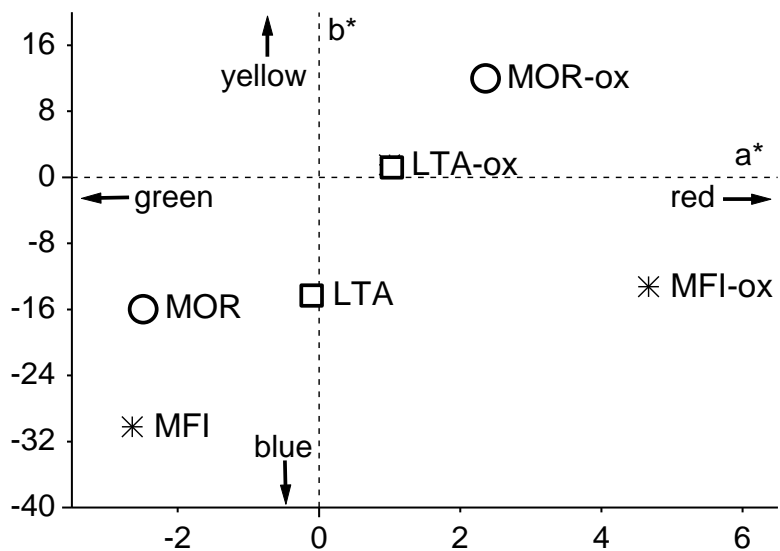

**Fig. 3** – Projection on the a*b* space of 3 indigo@zeolite samples before and after oxidising test (10 minutes in concentrated $HNO_3$, 14mol/l). LTA = indigo@LTA zeolite; MOR = indigo@mordenite; MFI = indigo@silicalite. Suffix –ox is added to distinguish samples after the oxidising test.

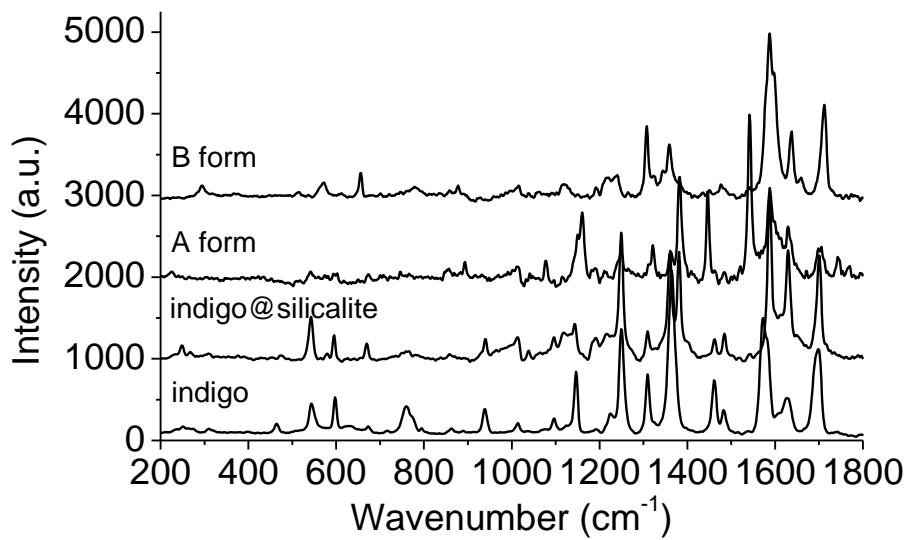

**Fig. 4** – Raman spectra of indigo powder, indigo@silicalite, A form (dehydroindigo) and B form (nitro-indigo). Spectra have been shifted for clarity after baseline correction (Laser power~100μW).

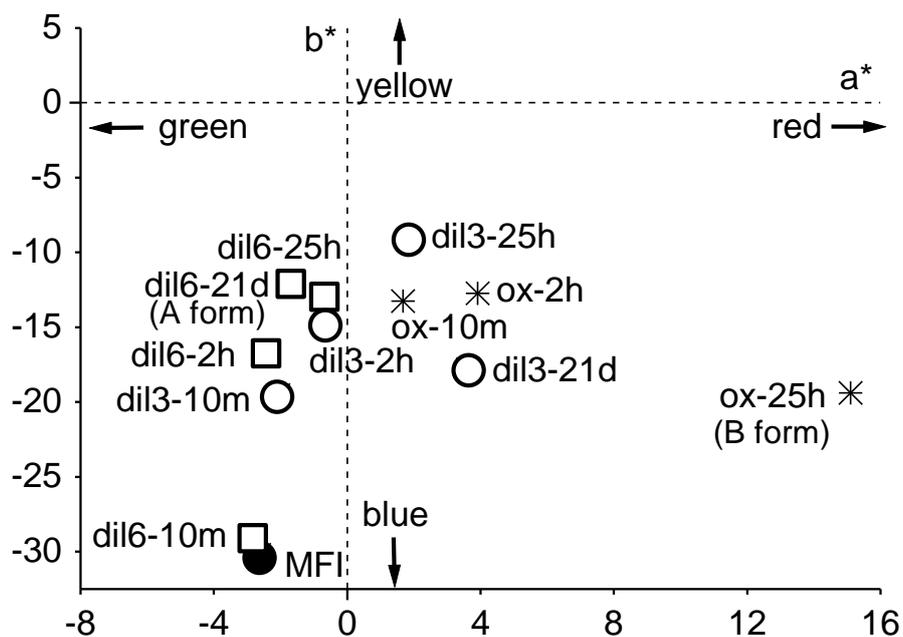

**Fig. 5 -** Projection on the a*b* space of several indigo@silicalite samples (●MFI refers to the indigo@silicalite sample before any test) after oxidising test with different $HNO_3$ concentrations (ox (∗) = concentrated $HNO_3$ (14mol/l); dil3 (○) = concentrated $HNO_3$ diluted 3 times; dil6 (□) = concentrated $HNO_3$ diluted 6 times) and duration time (10m = 10 minutes, 2h = 2 hours, 25h = 25 hours, 21d = 21 days).

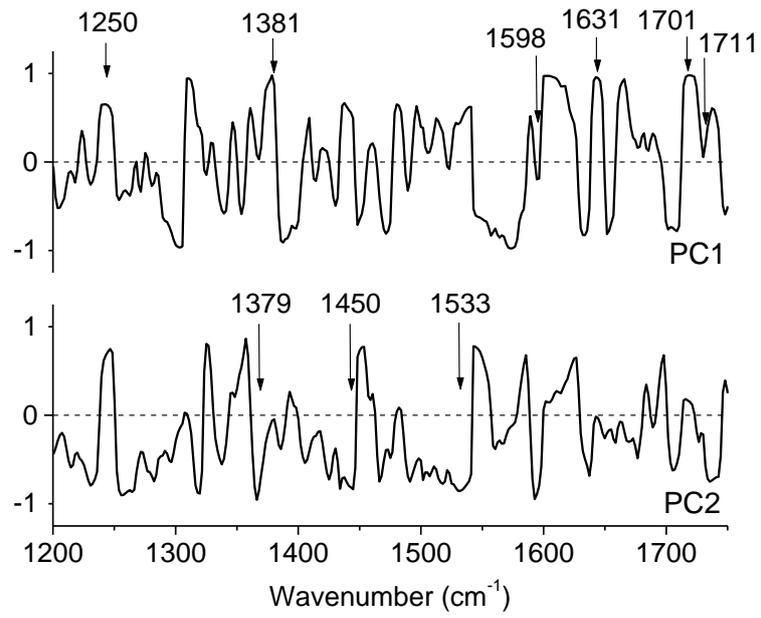

**Fig. 6 -** Loading plot of the two first Principal Components PC1 and PC2

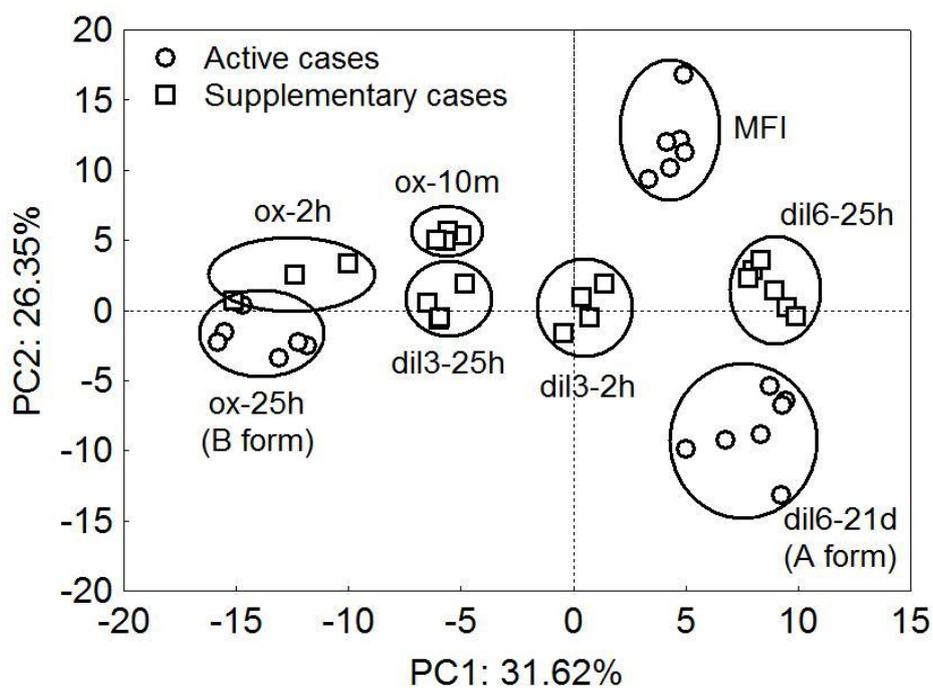

**Fig. 7** – Score plot for the first two principal components PC1 and PC2. MFI refers to the indigo@silicalite sample before any test); ox = concentrated $HNO_3$ (14mol/l), dil3 = concentrated $HNO_3$ diluted 3 times, dil6 = concentrated $HNO_3$ diluted 6 times; 10m = 10 minutes, 2h = 2 hours, 25h = 25 hours, 21d = 21 days.

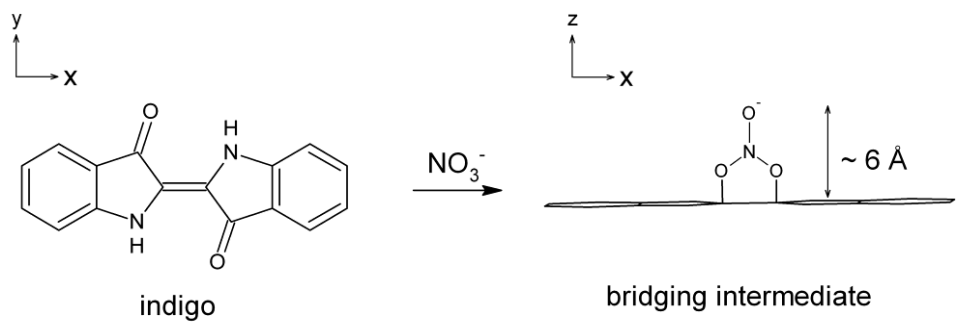

**Fig. 8** – Most probable geometric configuration between the indigo molecule and the $NO_3^-$ ion in the first step of the oxidisation process.

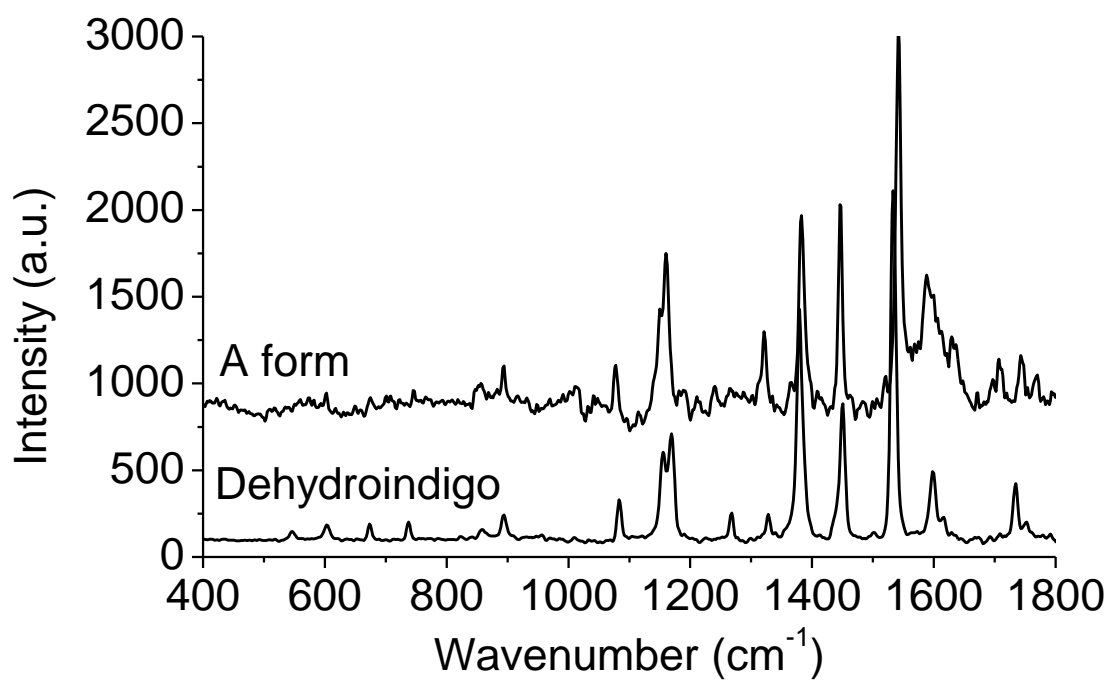

**Fig. 9** – Raman spectra of the freshly synthesized dehydroindigo and the A form. After baseline correction and scaling, the spectrum of indigo@silicalite sample in excess has been substracted from that of the A form presented on Fig. 4.

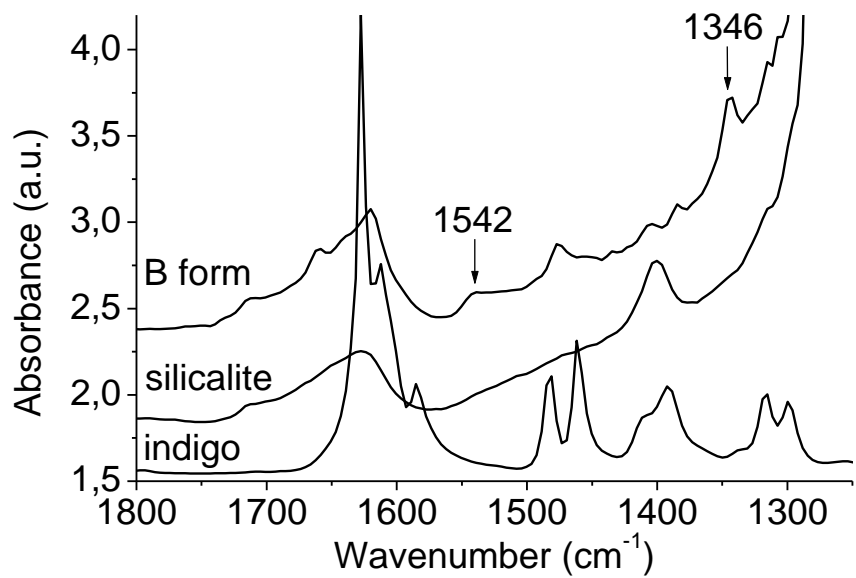

**Fig. 10 -** IR spectra of indigo, silicalite zeolite, and B form

**LIST OF FIGURES**

**Figure 1 -** Oxidisation of the indigo molecule

**Figure 2 –** Reflectance spectra of 3 indigo@zeolite samples (LTA = indigo@LTA zeolite; MOR = indigo@mordenite; MFI = indigo@silicalite), before the heating phase (a), b) and c)), after the heating phase (d), e) and f)), and after the oxidising test (10 minutes in concentrated $HNO_3$, 14mol/l, g), h) and i)). Suffixes –nh and –ox respectively refer to "non-heated" samples and samples after the oxidising test.

**Figure 3 –** Projection on the a*b* space of 3 indigo@zeolite samples before and after oxidising test (10 minutes in concentrated $HNO_3$, 14mol/l). LTA = indigo@LTA zeolite; MOR = indigo@mordenite; MFI = indigo@silicalite. Suffix –ox is added to distinguish samples after the oxidising test.

**Figure 4 –** Raman spectra of indigo powder, indigo@silicalite, A form (dehydroindigo) and B form (nitro-indigo). Spectra have been shifted for clarity after baseline correction (Laser power~100μW).

**Figure 5 -** Projection on the a*b* space of several indigo@silicalite samples (●MFI refers to the indigo@silicalite sample before any test) after oxidising test with different $HNO_3$ concentrations (ox (∗) = concentrated $HNO_3$ (14mol/l); dil3 (○) = concentrated $HNO_3$ diluted 3 times; dil6 (□) = concentrated $HNO_3$ diluted 6 times) and duration time (10m = 10 minutes, 2h = 2 hours, 25h = 25 hours, 21d = 21 days).

**Figure 6 -** Loading plot of the two first Principal Components PC1 and PC2

**Figure 7 –** Score plot for the first two principal components PC1 and PC2. MFI refers to the indigo@silicalite sample before any test); ox = concentrated $HNO_3$ (14mol/l), dil3 = concentrated $HNO_3$ diluted 3 times, dil6 = concentrated $HNO_3$ diluted 6 times; 10m = 10 minutes, 2h = 2 hours, 25h = 25 hours, 21d = 21 days.

**Figure 8 –** Most probable geometric configuration between the indigo molecule and the $NO_3^-$ ion in the first step of the oxidisation process.

**Figure 9 –** Raman spectra of the freshly synthesized dehydroindigo and the A form. After baseline correction and scaling, the spectrum of indigo@silicalite sample in excess has been substracted from that of the A form presented on Fig. 4.

**Figure 10 -** IR spectra of indigo, silicalite zeolite, and B form

# LIST OF TABLES

**Table 1** - Colour stability tests (10 minutes in concentrated nitric acid) for the different hybrid systems (1%wt. indigo) - Colorimetric indices are given in the La*b* space – LTA = indigo@LTA zeolite; MOR = indigo@mordenite; MFI = indigo@silicalite

**Table 2** - Oxidising tests for the indigo@silicalite (MFI) sample in presence of $HNO_3$ - Colorimetric indices are given in the La*b* space